# PB-AM: An Open-Source, Fully Analytical Linear Poisson-Boltzmann Solver

Lisa E. Felberg[1], David H. Brookes[2], Eng-Hui Yap[5], Elizabeth Jurrus[6,8],
Nathan Baker[7,9] Teresa Head-Gordon[1-4*]

[1]*Department of Chemical and Biomolecular Engineering,* [2]*Department of Chemistry,*
[3]*Department of Bioengineering, University of California Berkeley, Berkeley, California 94720*
[4]*Chemical Sciences Division, Lawrence Berkeley National Labs, Berkeley, California 94720*
[5]*Department of Systems and Computational Biology, Albert Einstein College of Medicine, Bronx, NY 10461*
[6] *Division of Computational and Statistical Analytics,* [7]*Advanced Computing, Mathematics, and Data Division, Pacific Northwest National Laboratory, Richland, Washington, 99352*
[8] *Scientific Computing and Imaging Institute, University of Utah, Salt Lake City, Utah, 84112*
[9] *Division of Applied Mathematics, Brown University, Providence, Rhode Island, 02912*

We present the open source distributed software package Poisson-Boltzmann Analytical Method (PB-AM), a fully analytical solution to the linearized Poisson Boltzmann equation, for molecules represented as non-overlapping spherical cavities. The PB-AM software package includes the generation of outputs files appropriate for visualization using VMD, a Brownian dynamics scheme that uses periodic boundary conditions to simulate dynamics, the ability to specify docking criteria, and offers two different kinetics schemes to evaluate biomolecular association rate constants. Given that PB-AM defines mutual polarization completely and accurately, it can be refactored as a many-body expansion to explore 2- and 3-body polarization. Additionally, the software has been integrated into the Adaptive Poisson-Boltzmann Solver (APBS) software package to make it more accessible to a larger group of scientists, educators and students that are more familiar with the APBS framework.

*Corresponding author
Email: thg@berkeley.edu

# 1. INTRODUCTION

The preferential association of biomolecules in the cell, ranging from proteins to nucleic acids to small metabolites, is crucial for cellular function. The first events in the molecular recognition process among multiple molecular constituents involve long-range diffusional association over large spatial scales. However, these dynamical encounters have proved difficult to model at an atomistic level since molecular dynamics is too computationally intensive to capture the large spatial and temporal scales over which many macromolecule associations occur. Fortunately coarse graining of the participating biomolecules and their environment, combined with simulations using stochastic dynamics, can be just as insightful as the all-atom deterministic dynamics. At these scales electrostatic interactions dominate, such that the coarse-graining strategy can rely on the popular Poisson-Boltzmann (PB) continuum mean field treatment that forms the basis of Gouy-Chapman theory[1,2] in electrochemistry, Derjaguin-Landau-Verwey-Overbeek (DLVO) theory in colloid chemistry[3,4], and under the low field linearized PB (LPB) approximation, the Debye-Hückel theory in solution chemistry[5].

There is a range of software packages currently available for solving the linearized Poisson-Boltzmann equation (LPBE). Numerical methods, such as finite-difference (FD) and finite-element (FE) packages including APBS[6], DelPhi[7], MEAD[8], UHBD[9], ZAP[10], and modules in Amber[11,12] and CHARMM[13,14], handle arbitrary dielectric boundaries by solving the PBE on a grid or mesh. Due to their numerical challenges, many essential features of the PBE were are not treated robustly in these software packages; e.g., the point charge singularity and the enforcement of electrostatic continuity across the dielectric boundary. Another limiting factor of these methods is the tradeoff between grid refinement and accuracy given the $O(m^3)$ scaling for 3D electrostatics, where m is the number of grid points in one dimension. Some of these issues have been mitigated for the energy: charge singularities have been treated with regularization schemes[15], induced surface charges[16] and other decomposition methods[17], molecular surface definitions and discontinuities across that boundary have been addressed with the matched interface and boundary (MIB) method[18] and advanced grid methods are also implemented[12,19]. However, unlike the energetic terms, force calculations still pose many numerical issues for FE and FD methods. The most definitive method, known as the 'virtual work' approach[20] has been implemented to account for full mutual polarization in FD and FE methods, but is limited by computational efficiency since gradients are calculated numerically, requiring a recalculation for the electrostatic energy after small displacements in each Cartesian direction.

Another approach is the boundary element method (BEM) that is the main feature of software packages such as AFMPB[21], PyGBe[22,23] and TABI[24]. BEM approaches offer some advantages over FE and FD methods since there is an analytical treatment of singular charges, and explicit treatment of the dielectric discontinuity and boundary conditions. However, BEMs also have some limitations, including an increasingly large dense matrix that scales with system size and has severe memory requirements, singular surface integrals, and issues with mesh generation. In contrast to FD and FE methods, BEMs can efficiently compute forces, but they only include self-polarization as opposed to full mutual polarization, and have yet to be demonstrated as efficient enough for dynamical simulations.

An alternative to the numerical FE, FD, and BEM methods is to consider analytical solutions instead. As far as we are aware, the first analytical LPBE solution was solved by Kirkwood for one spherical macromolecule.[25] However the treatment to describe mutual polarization requires at least two distinct spheres, and many different partial and approximate solutions to the mutual polarization of two or more macromolecules using the LPBE have been proposed.[26-30] In 2006 Lotan and Head-Gordon derived the first completely general analytical solution to the LPBE[31], including the forces and torques due to this potential, that permits analytical calculations to arbitrary large assemblies of interacting molecules. While the primary drawback of the analytical solutions are that they are restricted to idealized geometries such as spheres, they do have the benefit that the boundary conditions are solved completely, and for the Poisson-Boltzmann analytical model (PB-AM)[31] the mutual polarization is accounted for accurately.

While the PB-AM model[31] is computationally efficient, we believe that the model's theoretical formalism has made it inaccessible to many and inhibited its potential use for application on a large scale. Here we present an expanded, open-source software implementation of PB-AM, as well as a number of python-based utilities for creating files for multi-dimensional visualization for the electrostatic potential and forces and torques, a Brownian Dynamics simulation module that models the stochastic dynamics of multiple macromolecule interactions[32-34], the ability for users to define docking criteria and to simulate biomolecular association rate constants, as well as exploration of the many-body expansion (MBE) for the mutual polarization[35]. The PB-AM open source software is easy to download and install and has a simple input file structure and modular design to encourage both ease of use and extensibility to other software packages. Additionally, the PB-AM model has been integrated into the Adaptive Poisson-Boltzmann Solver (APBS) (http://www.poissonboltzmann.org/) to allow accessibility of the PB-AM capabilities to a larger group of scientists, educators and students.

The paper is organized as follows. After a brief review of the PB-AM theory developed by Lotan and Head-Gordon in Section 2, we describe the software framework of the PB-AM stand-alone code and its integration into the APBS software package in Section 3. In Section 4 we demonstrate the PB-AM software's visualization outputs for the electrostatic potential using 2D cross-sections as well as 3D isosurfaces that feed into the Visual Molecular Dynamics[36] (VMD) program. In Section 5 we introduce the dynamics capabilities using PB-AM that utilizes a standard Brownian dynamics scheme, permits users to define docking criteria, and offers two distinct ways to calculate bimolecular rate constants, one of which is especially suited to simulations involving large number of molecules. In Section 6 we introduce the many-body expansion (MBE) that allows the mutual polarization to be deconstructed into a direct polarization model, as well as approximate mutual polarization models. Finally we provide a brief summary in Section 7 and plans for future work.

## 2. PB-AM THEORY

The derivation details of PB-AM have been reported previously[31], but the main points can be summarized as follows. The electrostatic potential of the system at any point **r** is governed by the linearized form of the Poisson-Boltzmann equation

$$\nabla \cdot [\varepsilon(\boldsymbol{r})\nabla\Phi(\boldsymbol{r})] - \varepsilon(\boldsymbol{r})\kappa^2\Phi(\boldsymbol{r}) = 4\pi\rho(\boldsymbol{r}) \tag{1}$$

For the case of spherical cavities, we can solve Eq. (1) by dividing the system into inner sphere and outer sphere regions, and enforcing a set of boundary conditions that stipulate the continuity of the electrostatic potential and the electrostatic field at the surface of each sphere. The electrostatic potential inside molecule *i* is described by:

$$\Phi_{in}^{(i)}(\mathbf{t}) = \sum_{n=0}^{\infty}\sum_{m=-n}^{n}\left(\frac{E_{n,m}^{(i)}}{\varepsilon_p r^{n+1}} + B_{n,m}^{(i)} r^n\right) Y_{n,m}(\theta,\phi), \tag{2}$$

where $\varepsilon_p$ is the interior dielectric, $\mathbf{B}^{(i)}$ is a vector of unknowns that will be determined through the application of the boundary conditions, $\mathbf{E}^{(i)}$ is the multipole expansion of the charges inside molecule $i$.

$$E_{n,m}^{(i)} = \sum_{j=1}^{M_i} q_j^{(i)} \left(\rho_j^{(i)}\right)^n Y_{n,m}\left(\vartheta_j^{(i)}, \varphi_j^{(i)}\right). \tag{3}$$

where $M_i$ is the number of charges in molecule $i$, $q_j^{(i)}$ is the magnitude of the $j$-th charge, and $\mathbf{p}_j^{(i)} = \left[\rho_j^{(i)}, \alpha_j^{(i)}, \beta_j^{(i)}\right]$ is its position in spherical coordinates and $Y_{n,m}$ are the spherical harmonics. The general form of the potential outside all molecules (in a coordinate frame whose origin is the center of molecule $i$) is:

$$\Phi_{out}^{(i)}(t) = \frac{1}{\varepsilon_s} \sum_{n=0}^{\infty} \sum_{m=-n}^{n} \left( \frac{A_{n,m}^{(i)}}{r^{n+1}} e^{-\kappa r} \hat{k}_n(\kappa r) + L_{n,m}^{(i)} r^n \hat{i}_n(\kappa r) \right) Y_{n,m}(\theta, \phi), \tag{4}$$

where the coefficients $\mathbf{L}^{(i)}$ in Equation (4) are a re-expansion of the external potential coefficients $\mathbf{A}^{(j)}$, $j \neq i$ of all other molecules in the system. It is defined as:

$$\mathbf{L}^{(i)} = \sum_{\substack{j=1 \\ j \neq i}}^{N} \mathbf{T}^{(i,j)} \cdot \mathbf{A}^{(j)} \tag{5}$$

where $\mathbf{T}^{(i,j)}$ is the linear re-expansion operator that transforms a multipole expansion at $\mathbf{c}^{(j)}$ to a local (Taylor) expansion at $\mathbf{c}^{(i)}$. This operator is described in detail in our previous work[31]. The use of the $\mathbf{T}^{(i,j)}$ operators allows us to represent the potentials due to all molecules in the coordinate frame of a single molecule, a mathematical feature that is central to obtaining an analytical solution. Since the $\mathbf{B}^{(i)}$ depend on $\mathbf{A}^{(i)}$, the application of the boundary conditions leads to the following compact solution

$$\mathbf{A} = \Gamma \cdot (\Delta \cdot \mathbf{T} \cdot \mathbf{A} + \mathbf{E}) \tag{6}$$

Where $\mathbf{A}$ is a matrix of vectors, one for each molecule in the system, representing the effective multipole expansion of the charge distributions of each molecule, the $\Gamma$ matrix is a dielectric boundary-crossing operator, and the $\Delta$ matrix is a cavity polarization operator.

Using this formalism, physical properties of the system, such as interaction energy, forces and torques may be computed. The interaction energy of each molecule, $\Omega^{(i)}$, is the product of the molecule's total charge distribution (from fixed and polarization charges) with the potential due to external sources. This is computed as the inner product between the molecule's multipole expansion, $\mathbf{A}^{(i)}$, and the multipole expansions of the other molecules in the system, $\mathbf{L}^{(i)}$ as follows:

$$\Omega^{(i)} = \frac{1}{\epsilon_s} \langle \mathbf{L}^{(i)}, \mathbf{A}^{(i)} \rangle \qquad (7)$$

which allows us to define the force, which is computed as the gradient of the interaction energy with respect to the position of the center of molecule $i$ :

$$\mathbf{F}^{(i)} = -\nabla_i \Omega^{(i)} = \frac{-1}{\epsilon_s} \left( \langle \nabla_i \mathbf{L}^{(i)}, \mathbf{A}^{(i)} \rangle + \langle \mathbf{L}^{(i)}, \nabla_i \mathbf{A}^{(i)} \rangle \right) \qquad (8)$$

By definition, the torque on a charge in the molecule is the cross product of its position relative to the center of mass of the molecule with the force it experiences. The total torque on the molecule is a linear combination of the torque on all charges of the molecule.

$$\tau^{(i)} = \frac{1}{\epsilon_s} \left[ \mathbf{H}^{(i),x}, \mathbf{H}^{(i),y}, \mathbf{H}^{(i),z} \right] \times \left[ \nabla_i \mathbf{L}^{(i)} \right] \qquad (9)$$

Please see previous publication[31] for details on the PB-AM solver.

## 3. SOFTWARE ARCHITECTURE

The stand-alone PB-AM code has been completely re-written from the original code base development[31] in C++11 to include a modular interface, simplified input files, expanded examples, unit testing, and an automated build system (CMake). It also includes a set of utilities, written in Python, to aid in visualization tasks and perform additional simulations. The new Application Programming Interface (API) has been designed to focus on five important utilities of the program: 1) electrostatic potential visualization, 2) energy, force, and torque calculation for molecule-molecule interactions, 3) use of the many-body expansion (MBE) to formulate direct and mutual polarization models, 4) dynamical simulations using Brownian dynamics, and 5) different ways to calculate rate constants for biomolecule association. The input file structure is simple; each line containing a *keyword value* pair, in most cases, and for most basic usage, the program can be run with an input file of 3 lines. Additionally, the outputs have been expanded to formats readable in VMD[36], such as dx and xyz files. Finally, an extensive user manual and website (https://davas301.github.io/pb_solvers/) have been developed with examples and post-processing examples. PB-AM is available for download through the GitHub site: https://github.com/davas301/pb_solvers. Figure 1 shows the workflow of the software and its components.

In addition to the stand-alone code, the PB-AM code has been integrated into the Adaptive Poisson-Boltzmann Solver (APBS) software package that solves the equations of continuum electrostatics for large biomolecular assemblages. This software is the central computational element for computational research involving the Poisson-Boltzmann equation and is, a unique software package that solves the equations of continuum electrostatics for large biomolecular assemblages. APBS plays an important role in the structural and computational biology research community and the proposed research ensures its continued availability and support. Specifically, APBS addresses three key technology challenges for understanding solvation and electrostatics in biomedical applications: (1) accurate and efficient theories and models for biomolecular solvation and electrostatics, (2) robust and scalable software for applying those theories to biomolecular systems, and (3) mechanisms for sharing and analyzing biomolecular electrostatics data in the scientific community. PB-AM has been fully integrated into the APBS software. The user invokes the code with a keyword in the ELEC section of an APBS input file. Some additional keywords are required and are documented on the APBS website. The incorporation of PB-AM into APBS will allow for PB-AM to be available to an audience of thousands of users, and is made available through the APBS site: http://www.poissonboltzmann.org/.

## 4. ELECTROSTATIC POTENTIAL VISUALIZATION

One of the most popular uses of PBE solvers is to provide rapid analysis of the electrostatic potentials around macromolecules and their assemblies. Therefore, like many other PBE solvers, our PB-AM approach has the capability of producing output files for visualizing the electrostatic potential (ESP) of a system of arbitrary complexity. To this end we have provided several visualization options that a user may select, including two-dimensional ESP cross-sections and isosurface potentials (Figure 2), which may be loaded and viewed in the program Visual Molecular Dynamics[36] (VMD). In addition to the electrostatic potentials it is also quite straightforward to visualize of forces and torques on and around the center-of-mass.

## 5. DYNAMICAL TRAJECTORIES AND ANALYSIS TOOLS

In addition to electrostatics calculations, the PB-AM software includes dynamics capabilities. The computed expressions for force and torque have been incorporated into a Brownian dynamics simulation protocol adapted from Ermak and McCammon[32]. Each molecule in the system is treated as an independent rigid particle, and molecule-molecule overlap is not allowed. Assuming no hydrodynamic effects, the translational and rotational displacement, $\Delta \mathbf{r}_i$ and $\Delta \mathbf{\theta}_i$, respectively, are computed as:

$$\Delta r_i = \frac{D_{i,trans} \Delta t}{k_B T} f_i + S_i(\Delta t) \tag{10a}$$

$$\Delta \theta_i = \frac{D_{i,rot} \Delta t}{k_B T} \tau_i + \Theta_i(\Delta t) \tag{10b}$$

Where $S_i$ is the stochastic displacement, and $\Theta_i$ is the stochastic rotation, have the following properties in dimension $\alpha$ = x, y, z:

$$\langle S_\alpha \rangle = 0, \qquad \langle S_\alpha^2 \rangle = 2 D_{i_{trans}} \Delta t \tag{11a}$$

$$\langle \Theta_\alpha \rangle = 0, \qquad \langle \Theta_\alpha^2 \rangle = 2 D_{i_{rot}} \Delta t \tag{11b}$$

The PB-AM software allows a generalized simulation procedure for a variety of terminations conditions, including a) time: the simulation will terminate when it has run for t picoseconds, b) coordinates: the simulation will terminate when a specified particle has diffused beyond a certain point in space specified by the user, c) docking: the simulation will terminate when a pair of specified atoms in the simulated molecules are within a specified cutoff; this docking criteria is specified by the user as a contact list at the start of the simulation. For details on BD simulation in PB-AM, please see our user manual.

One application of dynamics of particular interest is the calculation of bimolecular association rate constants and the ability to specify the docking criteria when association is complete. For the case of PB-AM, which represents each molecule as one hard spherical object, the boundary of the sphere prevents the physical proximity of contact pairs to be defined based on an atomistic geometry. Instead, the locations of the atomistic contact points, from the contact list, are projected onto the surface of the sphere to create a reactive surface patch, and the BD simulation terminates at the docking criteria when these surface patches are within 0.1Å.

For calculations of the rate constants, we have incorporated into the PB-AM software package two independent approaches. The first is the Northrup-Allison-McCammon (NAM) method[37,38]. The NAM method is based on the Smoluchowski equation, and its underlying assumption is that the mobile molecule B only experiences forces from stationary molecule A; i.e., concentration effects that modulate intra-species interactions A:A and B:B are not taken into account. The NAM method models the rate calculation as an analysis across a series of flux surfaces: an inner region, bounded by a radial distance *b*, is defined where inter-molecular forces are anisotropic and must be evaluated explicitly with BD. In this region we can numerically evaluate the docking frequency, δ, i.e. the ratio of successfully docked trajectories to total trajectories simulated. For simulated distances outside *b*, any inter-molecular forces are approximately centrosymmetric and the diffusion rate constant $k_D(b)$ can be evaluated analytically using the Smoluchowski equation[37].

$$k(R) = \left[ \frac{1}{4\pi(D_{trans})} \int_R^\infty \frac{\exp(U(r)/k_B T)}{r^2} dr \right]^{-1} \tag{12}$$

Beyond the surface at *R=b*, a second surface at *R=q* is chosen such that $q >> b$, and the simulation can be truncated, and the diffusional rate beyond *q* can also be evaluated from the Smoluchowski equation, $k_D(q)$. By accounting for re-crossings across the two flux surfaces at *b* and *q*, we can evaluate intrinsic bimolecular collision rate, k, is as follows:

$$k = \frac{k_D(b)\delta}{1 - (1-\delta)k_D(b)/k_D(q)} \tag{13}$$

A more accurate rate constant calculation method must also simulate a system with multiple copies of A and B that are allowed to interact with each other. Given that such interactions are dominated by electrostatics at large distances, this demands an efficient algorithm capable of computing the electrostatic forces and torques for multiple molecules on the fly as in PB-AM. In this second approach it is possible to evaluate the mean first passage time (MFPT) under periodic boundary conditions, by considering the second-order rate equation of the association of molecules A and B:

$$\frac{d[AB]}{dt} = k[A][B] \tag{14}$$

In this case, the user can specify the concentrations [A] and [B] and the fraction of trajectories docked at time *t* is given by

$$P_{docked}(t) = \left( \text{No. of systems with } t_i \leq t \right)/N \tag{15}$$

At present one can use PB-AM to evaluate the pseudo-first order rate constant *k'* by fitting the plotted data of $P_{docked}(t)$ against *t*, to the functional form

$$P_{docked} = 1 - \exp\left(-k'(t - \tau_d)\right) \tag{16}$$

where $\tau_d$ is the dead time required for the system to equilibrate, and the bimolecular rate constant is computed from

$$k = k'/[A]_{t=0} \tag{17}$$

The barnase-barstar association kinetics has been extensively characterized, both experimentally[39] and computationally using the NAM method[40]. We compare the NAM method versus a multi-molecular simulation involving multiple barnase and barstar proteins ( ~ 125 proteins total). Both systems were performed under the following conditions: T = 298.15 K, dielectric constants of 78 (protein) and 4 (solvent), salt concentration of 0.05 M. A variable time step with a minimum of 2 ps was used. At each time step, the system is solved with a polarization cutoff of 10 Å and a force cutoff of 100 Å. The multi-molecular simulation was run with periodic boundary conditions with a box length of 320 Å to ensure that the minimum image conventions were obeyed. In Table 1 we compare these results against PB-AM using both NAM and the multi-protein MFPT simulations, showing that the simplified dielectric boundary is in very good qualitative agreement with past efforts, and thus a useful way to rapidly evaluate kinetics for biomolecule association. In comparison with the results of Gabdoulline and Wade, we believe that differences in computed rate constants are likely due to differences in molecular representations. All other system conditions held constant (protein model, force field, docking definition), the crowding conditions increase the docking rate by approximately ~20%.

## 6. MANY-BODY EXPANSION OF THE ELECTROSTATIC ENERGY

The formalism of the PB-AM model allows mutual polarization of all molecules in the system to be treated analytically, but as the number of molecules in the system increases, this N-body problem can become increasingly time intensive. The many-body expansion (MBE) allows us to expand the energy, and forces and torques as well, can be calculated in terms of simpler and independent 1-body, 2-body, 3-body etc. components.

$$U_N = U_1 + U_2 + U_3 + \cdots. \tag{18}$$

where

$$U_1 = \sum_{i=1}^{N} U(i) \; ; \; \Delta U_2 = \sum_{i=1}^{N-1} \sum_{j=i+1}^{N} U(i,j) - U(i) - U(j) \tag{19}$$

$$\Delta U_3 = \sum_{i=1}^{N-2}\sum_{j=i+1}^{N-1}\sum_{k=j+1}^{N} U(i,j,k) - U(i,j) - U(i,k) - U(j,k) + U(i) + U(j) + U(k) \quad (20)$$

We have shown that the MBE converges quickly at the 2-body level with small Debye lengths, whereas the 3-body truncation of Eq. (12) is excellent for larger Debye lengths.[35] This allows for a different type of refactoring of the calculations while still leading to a highly accurate approximation to the PB-AM model. The computational benefits to the approach are evident from Table 2, using a cubic grid of barnase and barstar molecules, in similar configurations to the MFPT initialization. We can achieve a 2-3X speed-up over the full-calculation using only 4 cores, and if we take advantage of the independent nature of the dimer and trimer calculations and increase the core count to 64, we can increase the speedup to 8X.

## 7. CONCLUSION

We have described the release of the PBE solver, PB-AM, a complete analytical solution for both self- and full mutual polarization, which allows users to analyze electrostatics and stochastic dynamics of complex biomolecular systems, under the assumption of idealized spherical geometries. PB-AM can be downloaded stand-alone or as part of the distributed APBS software package.

Obviously the simple geometric shape of the low dielectric spherical cavity containing the complex charge distribution can result in differences with that of a more detailed molecular shape handled by other LPBE formulations. We have overcome this limitation using our semi-analytic solution to the PBE (PB-SAM)[41,42] that expands the analytic formalism to describe molecular boundaries, and it will be made available in a future release of the PB-AM software package. Even so, we have demonstrated that meaningful rate constants can be evaluated under the idealized geometries inherent to the PB-AM model, as they were shown to be in good agreement with experiment and the PB-SAM solver that uses realistic geometric boundaries.

As a freely available, open-source package we hope that other users will find value in expanding the capabilities of the PB-AM code. In addition to the future release of PB-SAM, which describes more realistic molecular boundaries, there are other important additions that should be included such as hydrodynamic interactions and other kinetic schemes.

**ACKNOWLEDGEMENTS.**

THG and DHB were supported by the Director, Office of Science, Office of Basic Energy Sciences, of the U.S. Department of Energy under Contract No. DE-AC02-05CH11231. LF was supported by the National Science Foundation Graduate Research Fellowship under Grant No. DGE 110640. APBS is supported by the National Biomedical Computation Resource, the National Institutes of Health under grant GM069702. Work benefited from VMD software, under award National Institutes of Health grants 9P41GM104601 and 5R01GM098243-02, directed by Klaus Schulten. This research used resources of the National Energy Research Scientific Computing Center, a DOE Office of Science User Facility supported by the Office of Science of the U.S. Department of Energy under Contract No. DE-AC02-05CH11231. THG wishes Charles Brooks III the best on the occasion of his 60$^{th}$ birthday. Since I worked on the Poisson Boltzmann problem as one of the first projects as a graduate student in his group I think it fitting that the current status on PBE solvers be included in his Festschrift!

**TABLES**

**Table 1.** The barnase-barstar rate of association using different PBE models and rate constant protocols NAM and MFPT against experiment and using an atomistic docking criteria from Gabdoulline and Wade.

| Model and Rate Method | [Barnase] (concentration [M]):[Barstar] (concentration [M]) | Rate constant values, $k$ [ $M^{-1}s^{-1}$ ] |
|---|---|---|
| Experiment[39] | N/A | $2.86 \pm 0.28 \times 10^8$ |
| Gabdoulline and Wade[40] | 1 (5.068 x $10^{-5}$) : 1 (5.068 x $10^{-5}$) | $3.88 \pm 0. \times 10^8$ |
| PB-AM using NAM | 1 (5.068 x $10^{-5}$) : 1 (5.068 x $10^{-5}$) | $7.53 \pm 0.25 \times 10^7$ |
| PB-AM using MFPT | 124 (6.284 x $10^{-3}$) : 1 (5.068 x $10^{-5}$) | $9.58 \pm 0.47 \times 10^7$ |

**Table 2.** *Comparison of timings for full mutual polarization and the MBA.* The systems are comprised of a cubic grid of barnase and barstar molecules, with the given ratio of Barnase to Barstar molecules.

| | Timings (s) | | |
|---|---|---|---|
| [Barnase]:[Barstar] | Full Mutual (4 cores) | 3-body MBE (4 cores) | 3-body MBE (64 cores) |
| 7:1 | 1.07 | 3.19 | 0.47 |
| 63:1 | 376.22 | 185.11 | 50.81 |
| 124:1 | 2307.71 | 1612.34 | 297.01 |

**FIGURE CAPTIONS**

**Figure 1.** *Software workflow for the PB-AM model and its utilities.* The stand alone PB-AM code has been completely re-written from the original development[31] in C++11, and has four important utilities: electrostatic potential visualization, energy, force, and torque calculations for molecule-molecule interactions, use of the many-body expansion to formulate approximate direct and mutual polarization models, dynamical simulations using Brownian dynamics, which can generate many interesting outputs, for example, different ways to calculate rate constants for biomolecule association.

**Figure 2.** *Different visualizations of the electrostatic potentials from the PB-AM model.* All systems are at 0.0M salt concentration, 7 pH, with protonation states calculated using PROPKA[43], interior dielectric of 2, and solution dielectric of 78. (a) Three-dimensional surface potential from two different vantage points of the coarse-grained Barstar protein. The molecule has been coarse-grained into a single sphere of radius 21.8 Å that encompasses all molecule atoms, and the potential is depicted at the surface of the coarse-grain sphere. The color scale for each image pair is given by color bar on the right, in units of $k_BT/e$. (b) Three-dimensional isosurfaces for Omp32 Porin trimer from the view of (c) the exoplasmic face (channel entrance). The negative isosurface (red) forms a funnel that can direct anions from the environment towards the channel. The periplasmic face (channel exit). The positive surface (blue) at the channel exit may further enhance anion transport through the channel. The blue isosurface is drawn at 1.0 $k_BT/e$ and the red at -1.0 $k_BT/e$. The channel proteins are represented by a grey solvent-excluded surface. (d) The HIV glycoprotein binds to CD4 proteins (PDB structure 4NCO[44]) from three vantage points. The upper hemisphere is a virus membrane-spanning portion of the glycoprotein, while the lower hemisphere is connected to 3 CD4 proteins that extend from the surface of the cell wall. These two distinct binding regions of the protein are distinguished here by the location of the electrostatic potential isosurfaces, with the CD4 creating a positive potential, and the portion attached to virus membrane creating a negative potential. The blue isosurface is drawn at 1.0 $k_BT/e$ and the red at -1.0 $k_BT/e$.